# Minimal Neuron Circuits: Bursters

Amr Nabil, *Member, IEEE*, T. Nandha Kumar, and Haider Abbas F. Almurib, *Senior Member, IEEE*

*Abstract*— **This work introduces a novel methodology for designing biologically plausible bursting neuron circuits using a minimal number of components. We hypothesize that to design circuits capable of bursting, the neuron circuit design must mimic a neuron model that inherently exhibits bursting dynamics. Consequently, classical models such as the Hodgkin-Huxley, $I_{Na,p}+I_K$, and FitzHugh–Nagumo models are not suitable choices. Instead, we propose a methodology for designing neuron circuits that emulate the qualitative characteristics of the $I_{Na,p}+I_K+I_{K(M)}$ model, a well-established minimal bursting neuron model. Based on this methodology, we present two novel MOSFET-based circuits that exhibit bursting. Using the method of dissection of neural bursting, we demonstrate that the nullcline and bifurcation diagrams of the fast subsystem in our circuits are qualitatively equivalent to those of the $I_{Na,p}+I_K+I_{K(M)}$ model. Furthermore, we examine the effect of the type of bifurcation at burst initiation and termination on the bursting characteristics, showing that our circuits can exhibit diverse bursting behaviours. Importantly, the main contribution of this work lies not in the specific circuit implementation, but in the methodology proposed for constructing bursting neuron circuits.**

*Impact Statement* — **Commercial neuromorphic chips such as the Intel Loihi-2 and IBM TrueNorth offer approximately 1 million neurons per chip. These chips are typically utilized to construct large-scale neuromorphic systems such as the Intel Hala Point, which combines 1,152 Loihi-2 chips to achieve a total of around 1.15 billion neurons. This demand for extremely high neuron counts places stringent constraints on the circuit complexity and area of individual neurons. At the same time, advancing the performance and accuracy of spiking neural networks requires neurons with rich dynamics, such as bursting. To address these challenges, this work introduces a methodology for designing compact spiking neuron circuits that exhibit bursting. The proposed methodology aims to provide a flexible framework for circuit designers to tailor the neuron circuit designs to diverse system-level requirements.**

## I. INTRODUCTION

Spiking Neural Networks (SNNs) are an emerging class of Artificial Neural Networks designed to closely resemble the qualitative behaviours of Biological Neural Networks (BNNs). The advancement of SNNs is hindered by two major challenges. The first challenge lies in forming a complete understanding of the learning rules and mechanisms that govern processing and learning in BNNs. Once such understanding is mature, the second challenge becomes to design software or hardware that faithfully replicates these qualitative behaviours.

Mimicking the qualitative behaviours of BNNs typically requires capturing the qualitative dynamics of the fundamental building blocks of BNNs, namely the synapses and spiking neurons. Spiking neurons exhibit diverse behaviours that facilitate learning in both BNNs and SNNs, including homeostasis, lateral inhibition, subthreshold oscillations, and bursting. This work focuses primarily on the circuit-level implementation of neurons that demonstrate bursting.

Bursting is a neuronal firing pattern where a neuron fires a set of spikes referred to as a "burst", followed by a period of quiescence before the next burst. This phenomenon has been widely observed in biological neurons, including layers 3 and 5 pyramidal neurons [1], [2], cortical interneurons of the Neocortex [3], and pyramidal neurons in the $CA_1$ region of the Hippocampus [4]. Bursting is hypothesized to play a critical role in enhancing the reliability of communication between neurons [5]. Several mechanisms contribute to this improved reliability. For instance, a burst is more likely to evoke a post-synaptic spike than a single spike, as each spike within a burst generates excitatory post-synaptic potentials that accumulate. Additionally, bursts possess a larger signal-to-noise ratio than single spikes [6], owing to the threshold for bursting being larger than that of a single spike.

Furthermore, bursting can be used as a means of selective communication [7], where the presynaptic neuron targets a specific postsynaptic neuron. Such selective communication is achieved through resonance at either the synapse or neuron level. At the synaptic level, resonance emerges due to the interplay between short-term potentiation and depression [8], [9]. While at the neuronal level, resonance occurs when the frequency of presynaptic bursting matches the frequency of subthreshold oscillations at the postsynaptic neuron [10], [11], [12]. Since individual neurons can exhibit different subthreshold oscillation frequencies [13], [14], [15], bursting can provide a means of targeting specific postsynaptic neurons.

The $I_{Na,p}+I_K$ model discussed in [16], [17], [18] is a minimal model for spiking but not bursting. Consequently, it is incapable of demonstrating any bursting behaviour. However, if a very slow Potassium current (slower than the original $I_K$ current) is incorporated, the model can be extended to exhibit bursting. We denote this additional current throughout this work as $I_{K(M)}$. The addition of the $I_{K(M)}$ current to the $I_{Na,p}+I_K$ model results in the $I_{Na,p}+I_K+I_{K(M)}$ model, which is a minimal model for bursting. The aim of this work is to present a methodology for designing minimal bursting neuron circuits that mimic the qualitative characteristics of the $I_{Na,p}+I_K+I_{K(M)}$ model.

We begin by presenting the $I_{Na,p}+I_K+I_{K(M)}$ model in section II. Then, in section II-A, we discuss the dynamics of the $I_{Na,p}+I_K+I_{K(M)}$ model when the bursts are initiated through a saddle-node off an invariant circle bifurcation and terminated via a saddle homoclinic orbit bifurcation. Subsequently, in section II-B, we discuss the characteristics of the $I_{Na,p}+I_K+I_{K(M)}$ model with bursts that are initiated through a subcritical Andronov-Hopf bifurcation and terminated via a fold limit cycle bifurcation. Building on this foundation, sections III-A and III-B propose two novel MOSFET-based circuits that implement the $I_{Na,p}+I_K+I_{K(M)}$ model with bursts that are initiated and terminated through the aforementioned types of bifurcation. Finally, section IV compares the proposed circuits with existing neuron implementations reported in the literature.

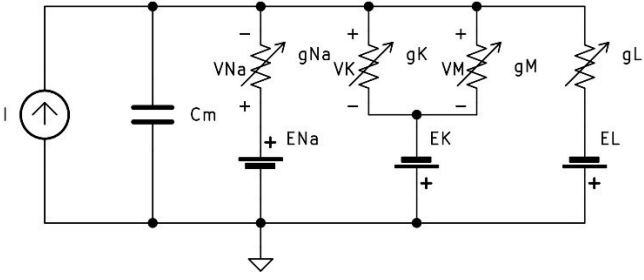

Fig. 1. The $I_{Na,p}+I_K+I_{K(M)}$ model circuit.

## II. THE $I_{Na,p}+I_K+I_{K(M)}$ MODEL

The $I_{Na,p}+I_K+I_{K(M)}$ model is a modification of the well-known $I_{Na,p}+I_K$ model. The additional $I_{K(M)}$ slow current modulates the model between bifurcations of the resting and spiking states. Therefore, this current is responsible for the termination and initiation of spiking between bursts. The $I_{Na,p}+I_K+I_{K(M)}$ model can be described using a set of three coupled ordinary differential equations.

$$C\frac{dV}{dt} = I - g_L(V - E_L) - g_{Na}m_\infty(V)(V - E_{Na}) -$$
$$g_K n(V - E_K) - g_M n_M(V - E_K) \qquad (1)$$

$$\frac{dn}{dt} = \frac{n_\infty(V) - n}{\tau(V)} \qquad (2)$$

$$\frac{dn_M}{dt} = \frac{n_{\infty,M}(V) - n_M}{\tau_M(V)} \qquad (3)$$

where $I$ is the injected current, and $C$ is the membrane capacitance. $E_K$, $E_{Na}$, and $E_L$ are voltage sources representing the Nernst potentials of the potassium, sodium, and leak ion channels, respectively. $g_K$, $g_M$, $g_{Na}$ and $g_L$ are the maximum conductance of the Potassium, slow Potassium, Sodium, and leak channels, respectively. $n$ and $n_M$ are normalized gating variables that control the flow of the Potassium current $I_K$ and the slow Potassium current $I_{K(M)}$. $\tau(V)$ and $\tau_M(V)$ are the time constants governing the rate of opening and closing of the gating variables $n$ and $n_M$. $m_\infty(V)$, $n_\infty(V)$ and $n_{\infty,M}(V)$ represent the steady state values of the gating variables as a function of the membrane potential, and can be approximated using a Boltzmann function given by

$$n_\infty(V) = \frac{1}{1 + e^{(V_{1/2} - V)/k}} \qquad (4)$$

where $V_{1/2}$ is the half-activation voltage satisfying $n_\infty(V_{1/2}) = 0.5$ and $k$ is the slope factor. The $I_{Na,p}+I_K+I_{K(M)}$ model can be divided into fast and slow subsystems, given by

$$\frac{dx}{dt} = f(x, u) \qquad (fast\ subsystem) \qquad (5)$$

$$\frac{du}{dt} = \mu g(x, u) \qquad (slow\ subsystem) \qquad (6)$$

where $x$ and $u$ are vectors representing the fast and slow variables, respectively. While $\mu$ represents the ratio of time scales between the fast and slow variables. For the $I_{Na,p}+I_K+I_{K(M)}$ model, the vector $x$ comprises the fast variables $V$ and $n$, whereas the vector $u$ consists of only the variable $n_M$. To analyse the mechanism of bursting in the $I_{Na,p}+I_K+I_{K(M)}$ model, we assume that $n_M = n_{\infty,M}(V)$. This assumption holds when $\tau_M \gg \tau$, ensuring $\mu$ is sufficiently small. Such an approach of studying bursting systems is known as "dissection" of neural bursting [19].

The fast subsystem here is equivalent to the $I_{Na,p}+I_K$ model. Therefore, the fast subsystem is capable of generating spiking independently of the slow subsystem. Nevertheless, for the system to exhibit bursting, the slow subsystem must modulate the fast subsystem to continuously alternate between a spiking and a resting state. In this work, we focus on discussing the mechanism by which the slow subsystem modulates the fast subsystem through bifurcations of the resting and spiking states. Therefore, we treat the variable $n_m$ as a bifurcation parameter in our analysis. Treating $n_M$ as a bifurcation parameter does not reveal bifurcations of the model as a whole, but only examines bifurcations of the fast subsystem. Nevertheless, examining the fast subsystem while varying $n_m$, provides insight into the neural characteristics of the system as a whole.

In the following sections, we study the fast subsystem of the $I_{Na,p}+I_K+I_{K(M)}$ model, and discuss the mechanism and bifurcations that lead to bursting. The objective is to establish a framework to evaluate the qualitative equivalence of the circuits proposed in section III. We discuss the types of bifurcation that lead to the disappearance of a resting state, which triggers the initiation of a burst. Likewise, we discuss the types of bifurcation that result in the disappearance of a stable limit cycle and lead to the termination of a burst. In two-dimensional systems, only four types of bifurcation can lead to a transition from a resting to a spiking state, and only four types can lead to a transition from a spiking to a resting state. Therefore, 16 combinations of those types of bifurcation can lead to bursting. However, we only discuss two of those combinations in the following sections.

### A. Saddle-node off Invariant Circle / Saddle Homoclinic Orbit Bifurcation

Fig. 2 shows the membrane potential output of the $I_{Na,p}+I_K+I_{K(M)}$ model in response to a fixed injected current. Fig. 2 (a) shows that at I = 4 mA, no spiking or bursting behaviour is observed and the membrane potential converges to a resting value. In contrast, if the injected current is increased to I = 5 mA, the model outputs sustained bursting of the membrane potential as shown in Fig. 2 (b). The bursting observed in Fig. 2 (b) is attributed to a stable limit cycle in the whole three-dimensional system. However, when we dissect the system and examine only the two-dimensional fast subsystem, the bursting corresponds to the membrane potential continuously alternating between a stable equilibrium and a stable limit cycle. In this case, $n_M$ is responsible for the transition from the resting to the spiking state and vice versa.

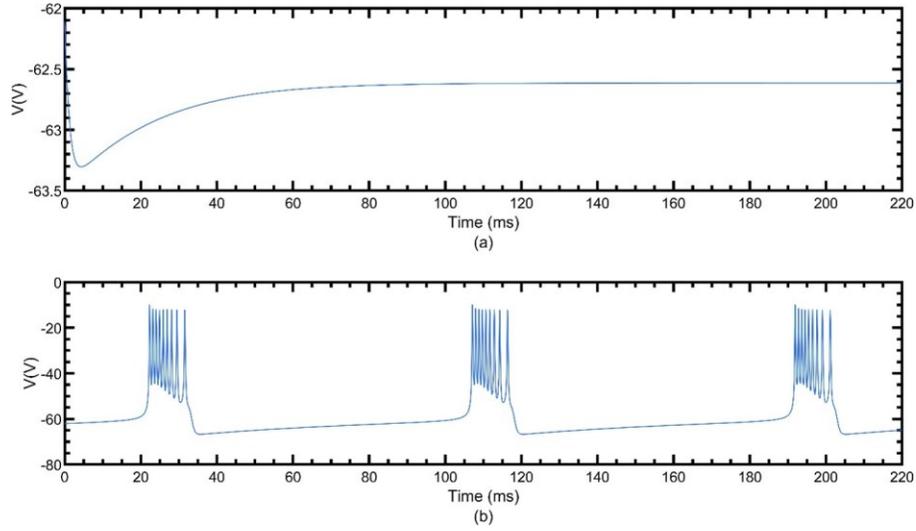

Fig. 2. Membrane potential output of the $I_{Na,p}+I_K+I_{K(M)}$ model with $C = 1\ mF$, $E_L = -80\ mV$, $E_{Na} = 60\ mV$, $E_K = -90\ mV$, $g_L = 8\ S$, $g_{Na} = 20\ S$, $g_K = 9\ S$, $V_{1/2Na} = -20\ mV$, $V_{1/2K} = -25\ mV$, $V_{1/2M} = -20\ mV$, $k_{Na} = 15*10^{-3}$, $k_K = 5*10^{-3}$, $K_M = 5*10^{-3}$, $\tau = 0.152\ ms$ and $\tau_M = 20\ ms$. (a) I = 4 mA. (b) I = 5 mA. [16]

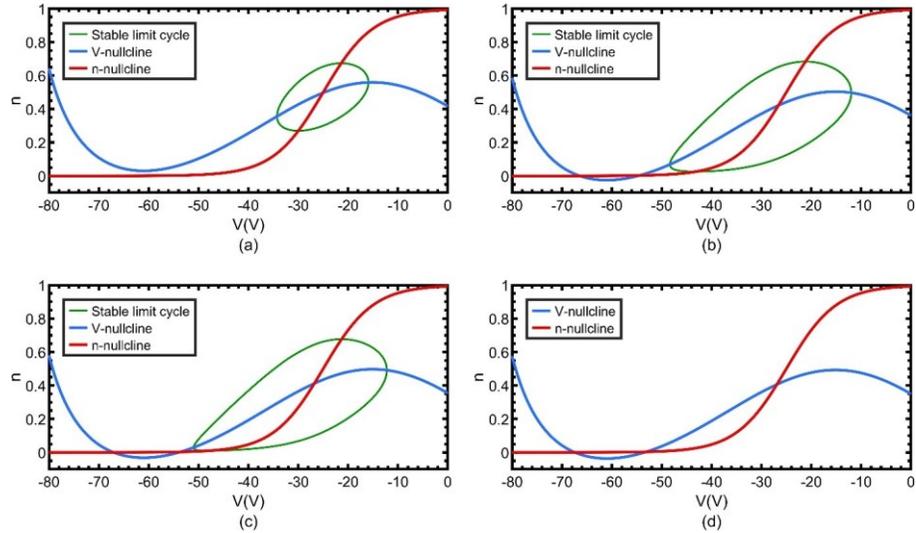

Fig. 3. Nullcline analysis of the $I_{Na,p}+I_K+I_{K(M)}$ model with the same parameters values as Fig. 2 (b). (a) $n_M$=-0.05. (b) $n_M$=0.05. (c) $n_M$=0.062. (d) $n_M$=0.07.

Fig. 3 illustrates the nullclines of the fast subsystem for different values of the gating variable $n_M$. Before the burst is initiated, the membrane potential of the fast subsystem would be at the resting value of a stable equilibrium. This stage is demonstrated using Fig. 3 (b), which shows that at $n_M = 0.05$, the fast subsystem has three equilibria and a stable limit cycle. The leftmost equilibrium is a stable node, the middle is a saddle, and the rightmost is an unstable focus. At this stage, the membrane potential of the system resides at the stable node. Nevertheless, $n_M$ decreases during the quiescent phase, which shifts the v-nullcline upwards until the stable node and saddle coalesce. As $n_M$ decreases further, the stable node and saddle disappear via a saddle-node off an invariant circle bifurcation as shown in Fig. 3 (a). Consequently, the membrane potential converges to the stable limit cycle, and a burst is initiated.

During the burst, $n_M$ increases and the stable node and saddle equilibria reappear through the saddle-node bifurcation. Despite this, the membrane potential remains at the stable limit cycle. As $n_M$ increases further, the stable limit cycle approaches the saddle as shown in Fig. 3 (c). Once the limit cycle touches the saddle, it forms a homoclinic trajectory around the saddle. Therefore, as $n_M$ increases even further, the limit cycle disappears through a saddle homoclinic orbit bifurcation as shown in Fig. 3 (d). Consequently, the membrane potential converges to the resting value of the stable node, and the burst is terminated. During the subsequent quiescent phase, $n_M$ decreases again, and the stable limit cycle reappears through the saddle homoclinic orbit bifurcation. However, the membrane potential remains at the quiescent phase until $n_M$ decreases further, and the stable node disappears via the saddle-node bifurcation. This cycle results in the sustained bursting shown in Fig. 2 (b).

Fig. 4 (a) shows the bifurcation diagram of the fast subsystem with $n_M$ as the bifurcation parameter. The bifurcation diagram shows that two bifurcations occur as $n_M$ is varied. A saddle-node bifurcation occurs at $n_M \approx 0.01$, which results in the appearance or disappearance of a stable node and a saddle. Additionally, a saddle homoclinic orbit bifurcation occurs at $n_M \approx 0.065$, which results in the appearance or disappearance of a stable limit cycle.

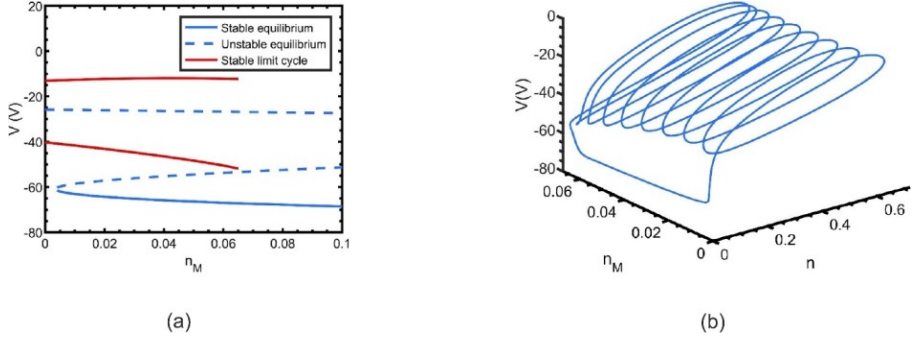

Fig. 4. (a) Bifurcation diagram and (b) trajectory of membrane potential bursting of the $I_{Na,p}+I_K+I_{K(M)}$ model. The same parameter values in Fig. 2 (b) were used.

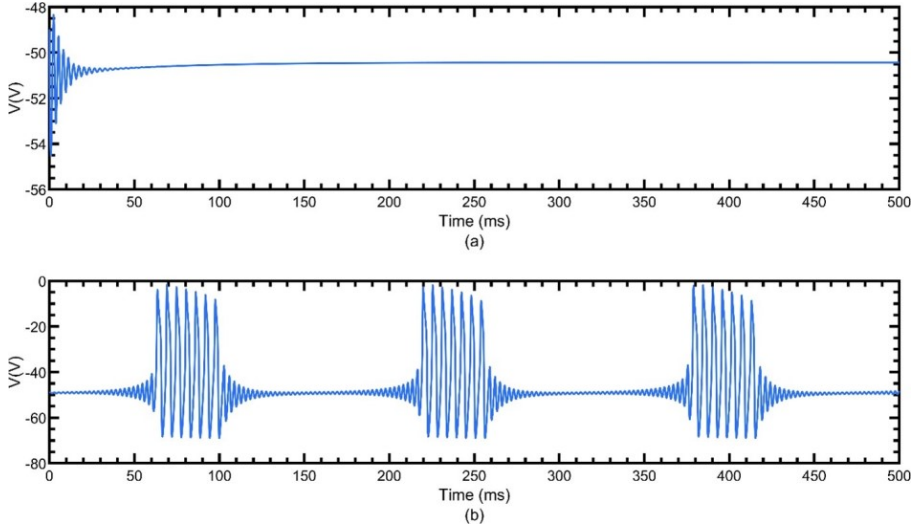

Fig. 5. Membrane potential output of the $I_{Na,p}+I_K+I_{K(M)}$ model with $C = 1\ mF$, $E_L = -78\ mV$, $E_{Na} = 60\ mV$, $E_K = -90\ mV$, $g_L = 1\ S$, $g_{Na} = 4\ S$, $g_K = 4\ S$, $V_{1/2_{Na}} = -30\ mV$, $V_{1/2_K} = -45\ mV$, $V_{1/2_M} = -20\ mV$, $k_{Na} = 7*10^{-3}$, $k_K = 5*10^{-3}$, $K_M = 5*10^{-3}$, $\tau = 1\ ms$ and $\tau_M = 60\ ms$. (a) I = 45 mA. (b) I = 55 mA. [16]

The types of bifurcation that initiate and terminate the burst affect the characteristics of the resulting bursting output [16]. Since the bursting in this section is initiated through a saddle-node bifurcation, the spikes at the beginning of the burst are all-or-nothing spikes with no subthreshold oscillation, as shown in Fig. 2 (b). Similarly, since the bursting is terminated via a saddle homoclinic orbit bifurcation, no subthreshold oscillations are present towards the end of the burst. This can also be inferred from the three-dimensional trajectory of the burst shown in Fig. 4 (b). Because different bifurcation types govern burst initiation and termination, the neuronal characteristics at the beginning and end of a burst can differ. For instance, subthreshold oscillations might exist at the start of the burst but not at the end, and vice versa.

### B. Subcritical Andronov-Hopf / Fold Limit Cycle Bifurcation

In this section, we modify the model's parameters so that each burst is initiated through a subcritical Andronov-Hopf bifurcation and terminated via a fold limit cycle bifurcation. Fig. 5 (a) shows that the model's membrane potential converges to a resting state at $I = 45\ mA$. But if the injected current is increased to $I = 55\ mA$, the membrane potential shows sustained bursting as illustrated in Fig. 5 (b). Similar to the previous section, the sustained bursting shown in Fig. 5 (b) corresponds to a stable limit cycle in the whole three-dimensional system.

Fig. 6 shows the nullclines of the fast subsystem of the model with varying $n_M$. At the quiescent phase of bursting, the system has a stable focus equilibrium surrounded by an unstable limit cycle, which is encircled by another stable limit cycle as shown in Fig. 6 (b) and Fig. 6 (c). At this phase, the membrane potential resides at the stable focus, causing $n_M$ to decrease and the unstable limit cycle to shrink as depicted in Fig. 6 (b). As $n_M$ decreases further, the unstable limit cycle shrinks and disappears. Consequently, the stable focus loses stability through a subcritical Andronov-Hopf bifurcation, which is illustrated using Fig. 6 (a). Once the equilibrium becomes unstable, the membrane potential converges to the stable limit cycle, and a burst is initiated.

During the spiking phase, $n_M$ begins to increase. Eventually, the unstable equilibrium reappears, and the equilibrium regains stability through the same subcritical Andronov-Hopf bifurcation. Despite this, the membrane potential remains at the stable limit cycle. As $n_M$ increases, the unstable limit cycle expands and approaches the stable limit cycle as shown in Fig. 6 (c). $n_M$ continues to increase until the unstable and stable limit cycles collide and annihilate via a fold limit cycle bifurcation as illustrated in Fig. 6 (d). Consequently, the membrane potential converges to the stable equilibrium, and the burst is terminated. During the following quiescent phase, $n_M$ starts to decrease again, causing the stable and unstable limit cycles to reappear through the same fold limit cycle bifurcation. Nevertheless, the membrane potential remains at the quiescent phase until the equilibrium loses stability again through the subcritical Andronov-Hopf bifurcation. This process underlies the sustained bursting observed in Fig. 5 (b).

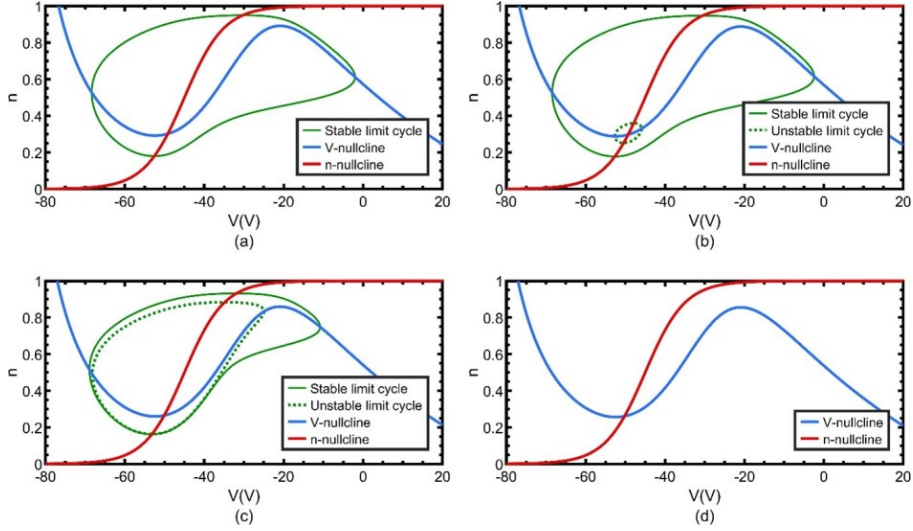

Fig. 6. Nullcline analysis of the $I_{Na,p}+I_K+I_{K(M)}$ model with the same parameters values as Fig. 5 (b). (a) $n_M$=0.055. (b) $n_M$=0.065. (c) $n_M$=0.14. (d) $n_M$=0.15.

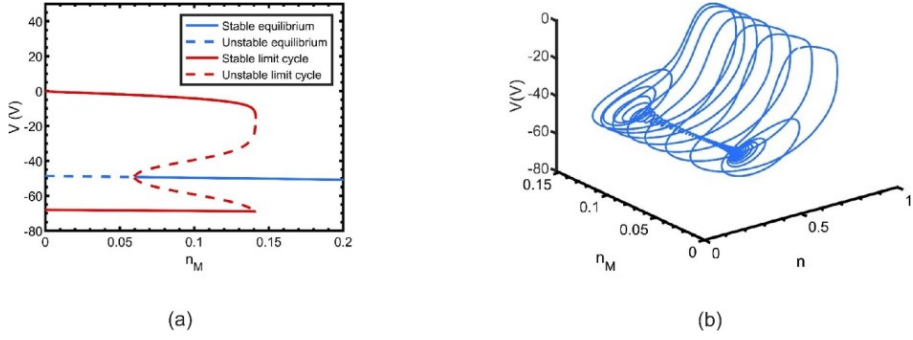

Fig. 7. (a) Bifurcation diagram and (b) trajectory of membrane potential bursting of the $I_{Na,p}+I_K+I_{K(M)}$ model. The same parameter values in Fig. 5 (b) were used.

Fig. 7 (a) presents the bifurcation diagram of the fast subsystem as a function of $n_M$. As in the previous case, two bifurcations are observed. The subcritical Andronov-Hopf bifurcation at $n_M \approx 0.06$ is responsible for the equilibrium losing stability and initiating the burst. While the fold limit cycle bifurcation at $n_M \approx 1.142$ is responsible for the disappearance of the stable limit cycle and the termination of the burst. Since the bifurcation types in this case differ from those in the previous section, the resulting bursting exhibits distinct qualitative characteristics. Specifically, the bursting membrane potential waveform in Fig. 5 (b) shows subthreshold oscillations at the start and end of each cycle, in contrast to the all-or-nothing spikes seen in Fig. 2 (b). The subthreshold oscillations are also clearly observed in the three-dimensional trajectory in Fig. 7 (b).

## III. FET-based Minimal Bursting Neuron

This section presents a methodology for designing minimal spiking neuron circuits that are capable of bursting. The circuits proposed in this section qualitatively agree with the $I_{Na,p}+I_K+I_{K(M)}$ model, whilst using a minimal number of components. It has been shown that a circuit comprising a Type-N Negative Differential Resistance (NNDR) device or circuit, along with a MOSFET and RC circuit exhibits neuronal characteristics that are qualitatively similar to the $I_{Na,p}+I_K$ model [18]. To design minimal neuron circuits that imitate the $I_{Na,p}+I_K+I_{K(M)}$ model instead, we only need to add circuitry representing the additional $I_{K(M)}$ current.

To demonstrate the qualitative similarity between our circuits and the $I_{Na,p}+I_K+I_{K(M)}$ model, we adopt the same method of dissection of neural bursting throughout this section. As with the $I_{Na,p}+I_K+I_{K(M)}$ model, 16 combinations of different types of bifurcation can result in bursting in our circuits, though we illustrate only two representative cases here. In section III-A, we present a novel neuron circuit that produces bursting initiated by a saddle-node off an invariant circle bifurcation and terminated via a saddle homoclinic orbit bifurcation. While in section III-B, we modify the circuit's parameters so that the circuit outputs a bursting waveform, where each burst is initiated through a subcritical Andronov-Hopf bifurcation and terminated through a fold limit cycle bifurcation.

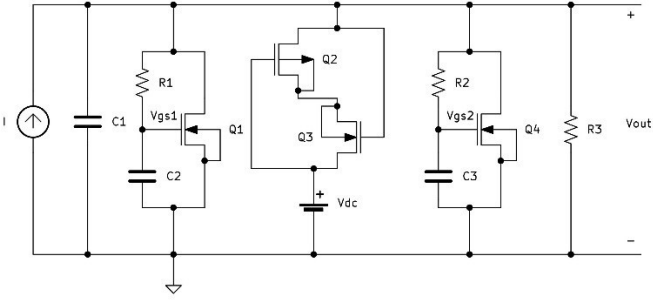

Fig. 8. FET-based minimal bursting neuron circuit. (Saddle-node off Invariant Circle / Saddle Homoclinic Orbit bifurcation)

### A. Saddle-node off Invariant Circle / Saddle Homoclinic Orbit Bifurcation

We propose a novel MOSFET-based minimal neuron circuit capable of bursting in Fig. 8. This design builds on the circuits presented in [18], with the addition of a slow $I_{K(M)}$ circuitry. Since both $I_K$ and $I_{K(M)}$ are Potassium currents, the slow $I_{K(M)}$ circuitry is the same MOSFET and RC circuitry used to represent the $I_K$ current in [18]. The only modification is that the RC time constant is chosen to be significantly larger. To analyse the circuit, we first derive the circuit's equations.

$$\frac{dV_{out}}{dt} = \frac{1}{C_1}(I - \frac{V_{out} - V_{GS1}}{R_1} - i_{DSQ_1} + i_{DSQ_3} -$$
$$\frac{V_{out} - V_{GS2}}{R_2} - i_{DSQ_4} - \frac{V_{out}}{R_3}) \tag{7}$$

$$\frac{dV_{GS1}}{dt} = \frac{1}{C_2 R_1}(V_{out} - V_{GS1}) \tag{8}$$

$$\frac{dV_{GS2}}{dt} = \frac{1}{C_3 R_2}(V_{out} - V_{GS2}) \tag{9}$$

where $V_{out}$, $V_{GS1}$ and $V_{GS2}$ are the state variables of the circuit, and represent the voltages across the capacitors $C_1$, $C_2$ and $C_3$, respectively. Similar to the circuits discussed in [18], $Q_1$, $R_1$ and $C_2$ represent the $I_K$ current, while the NNDR circuit comprising $Q_2$ and $Q_3$ represents the $I_{Na}$ current, and $R_3$ represents the $I_L$ current. The additional circuitry comprising $Q_4$, $R_2$ and $C_3$ represent the additional $I_{K(M)}$ current. We have discussed in section II the mechanism by which the $I_{K(M)}$ current drives the model through bifurcations of the spiking and resting states. Here, the current through the MOSFET $Q_4$ fulfils a similar role. Additionally, the voltage $V_{GS2}$ governing the opening and closing of $Q_4$'s channel fulfils the same role as the gating variable $n_M$.

Fig. 9 (a) illustrates that the membrane potential of the circuit in Fig. 8 converges to a resting state and does not show spiking or bursting at $I = 0.8 \ uA$. However, when the injected current is increased to $I = 1.2 \ uA$, the circuit outputs sustained bursting as shown in Fig. 9 (b). This bursting is due to a bifurcation that results in the appearance of a stable limit cycle in the three-dimensional system. To analyse the underlying mechanism of this bursting, we divide the circuit into a fast subsystem comprised of the state variables $V_{out}$ and $V_{GS1}$, and a slow subsystem comprised of the state variable $V_{GS2}$. Similar to the $I_{Na,p}+I_K+I_{K(M)}$ model, when bursting occurs, the slow subsystem drives the fast subsystem into bifurcations that result in the initiation and termination of spiking. To investigate these bifurcations, $V_{GS2}$ is treated as the bifurcation parameter.

Fig. 10 shows the nullclines of the circuit's fast subsystem for different values of $V_{GS2}$. These nullclines are obtained by equating the derivatives $dV_{out}/dt$ and $dV_{GS1}/dt$ to zero and solving the equations numerically. To explain the mechanism of bursting in the circuit of Fig. 8, we first assume that the circuit is at the quiescent phase of bursting. Fig. 10 (b) shows the nullclines of the circuit at $V_{GS2} = 1.16 \ V$, where three equilibria and a stable limit cycle exist. The leftmost equilibrium is a stable node, the middle is a saddle, and the rightmost is an unstable focus. During the quiescent phase, the membrane potential resides at the left stable node. Since the resting value of the stable node is smaller than $V_{GS2}$, the capacitor $C_3$ starts discharging, and $V_{GS2}$ decreases towards $V_{out}$. The decrease in $V_{GS2}$ causes the V-nullcline to shift downwards until the stable node and saddle coalesce then annihilate via a saddle-node off an invariant circle bifurcation. Therefore, the stable equilibrium and the saddle disappear as shown in Fig. 10 (a). Consequently, the membrane potential converges to the stable limit cycle, and a burst is initiated.

During the burst, $V_{out}$ becomes larger than $V_{GS2}$. Therefore, the capacitor $C_3$ starts charging and $V_{GS2}$ increases. The increase in $V_{GS2}$ causes the stable node and saddle to reappear via the same saddle-node bifurcation as shown in Fig. 10 (b). Nevertheless, the membrane potential remains at the stable limit cycle. As $V_{GS2}$ increases, the stable limit cycle approaches the saddle as shown in Fig. 10 (c). Eventually, the stable limit cycle forms a homoclinic trajectory around the saddle, and as $V_{GS2}$ increases further, the stable limit cycle disappears via a saddle homoclinic orbit bifurcation as illustrated in Fig. 10 (d). Once the limit cycle disappears, the membrane potential converges to the stable node, and the burst is terminated.

Consequently, the capacitor $C_3$ starts discharging and $V_{GS2}$ decreases once again. Therefore, the stable limit cycle reappears through the same saddle homoclinic orbit bifurcation as shown in Fig. 10 (c). However, the membrane potential remains at the stable node until the saddle-node bifurcation occurs and the equilibria disappear. Once the equilibria disappear as shown in Fig. 10 (a), the membrane potential converges again to the stable limit cycle, and another burst is initiated. This process results in the sustained bursting shown in Fig. 9 (b) and is qualitatively equivalent to that of the $I_{Na,p}+I_K+I_{K(M)}$ model discussed in section II-A.

Fig. 11 (a) shows the bifurcation diagram of the circuit's fast subsystem with $V_{GS2}$ as the bifurcation parameter. It can be inferred from Fig. 11 (a) that two bifurcations occur as $V_{GS2}$ is varied. The saddle-node bifurcation at $V_{GS2} \approx 1.14 \ V$ results in the disappearance of the stable node and the initiation of the burst. While the saddle homoclinic orbit bifurcation at $V_{GS2} \approx 1.228 \ V$ is responsible for the disappearance of the stable limit cycle and the termination of the burst. For $V_{GS2} < 1.14 \ V$, the fast subsystem does not have any stable equilibria. Thus, the membrane potential always converges to the stable limit cycle, and the circuit would be at the spiking phase of bursting. On the other hand, for $V_{GS2} > 1.228 \ V$, the fast subsystem does not have any stable limit cycles. Therefore, the membrane potential always converges to the resting value of the stable node, and the circuit would become at the quiescent phase of bursting. However, the fast subsystem is bistable for $1.14 < V_{GS2} < 1.228$, which results in a hysteresis effect. This means that for

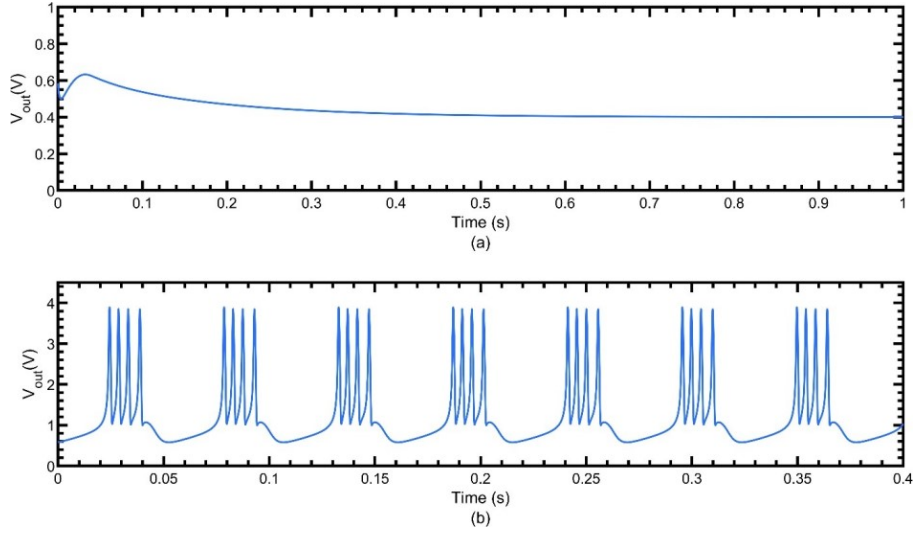

Fig. 9. Membrane potential output of the FET-based minimal bursting neuron circuit in Fig. 8 with $C_1 = 5\ nF$, $C_2 = 1\ nF$, $R_1 = 1\ M\Omega$, $V_{dc} = 4.5\ V$, $K_{n1} = 100\ uA/V^2$, $V_{t01} = 2\ V$, $\lambda_1 = 0.01$, $K_{p2} = 40\ uA/V^2$, $V_{t02} = 2\ V$, $\lambda_2 = 0.01$, $K_{n3} = 40\ uA/V^2$, $V_{t03} = -2\ V$, $\lambda_3 = 0.01$, $C_3 = 100\ nF$, $R_2 = 1\ M\Omega$, $K_{n4} = 40\ uA/V^2$, $V_{t04} = 1\ V$, $\lambda_4 = 0.01$ and $R_3 = 0.5\ M\Omega$. (a) I = 0.8 uA. (b) I = 1.2 uA.

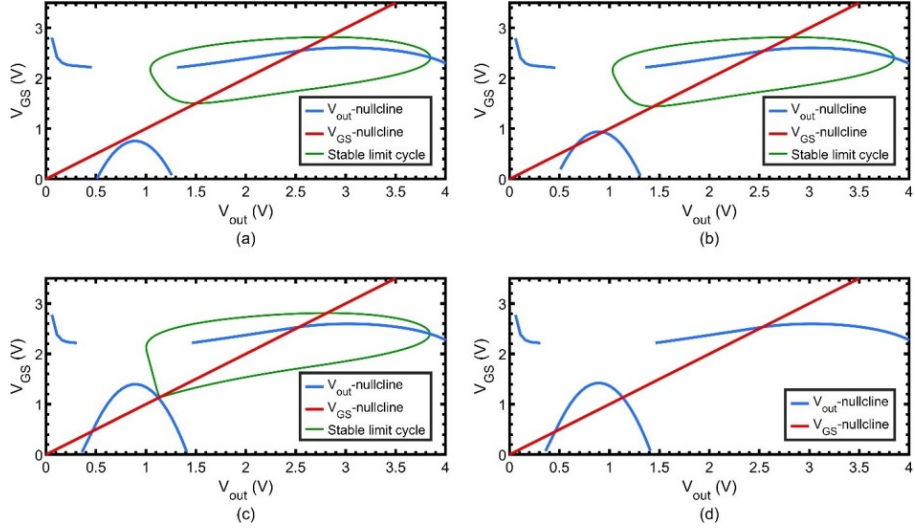

Fig. 10. Nullcline analysis of the FET-based minimal bursting neuron circuit in Fig. 8 with the same parameters values as Fig. 9 (b). (a) $V_{gs2}$=1.12 V. (b) $V_{gs2}$=1.16 V. (c) $V_{gs2}$=1.227 V. (d) $V_{gs2}$=1.23 V.

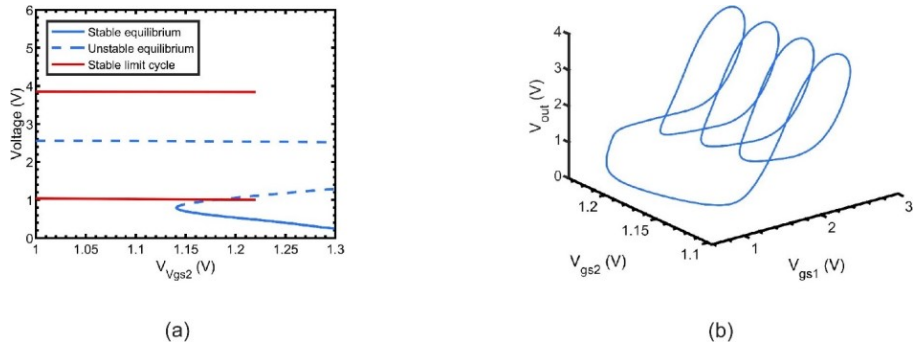

Fig. 11. (a) Bifurcation diagram and (b) trajectory of membrane potential bursting of the FET-based minimal bursting neuron circuit in Fig. 8. The same parameters values in Fig. 9 (b) were used.

$1.14 < V_{GS2} < 1.228$, the circuit could be at either the quiescent or spiking phases, depending on the state of the circuit. The bifurcation diagram in Fig. 11 (a) is qualitatively equivalent to that of the $I_{Na,p}$+$I_K$+$I_{K(M)}$ model discussed in section II-A, which highlights the qualitative similarity between our circuit and the $I_{Na,p}$+$I_K$+$I_{K(M)}$ model.

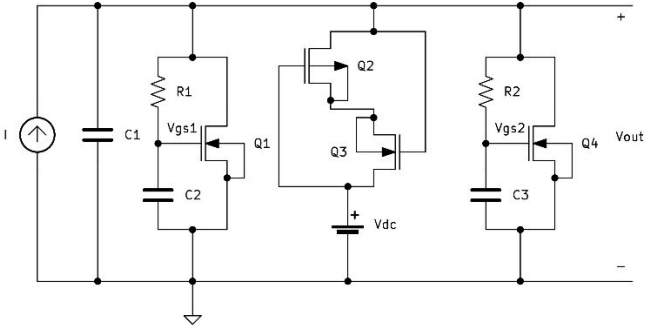

Fig. 12. FET-based minimal bursting neuron circuit. (Subcritical Andronov-Hopf / Fold Limit Cycle bifurcation)

As discussed in section II for the $I_{Na,p}+I_K+I_{K(M)}$ model, the types of bifurcation governing the initiation and termination of a burst affect the burst's characteristics. The same principle applies to the circuit in Fig. 8. The bursting shown in Fig. 9 (b) comprises all-or-nothing spikes with no subthreshold oscillations. The absence of subthreshold oscillations at the onset of each burst reflects the nature of the saddle-node bifurcation. Likewise, the lack of subthreshold oscillations at the end of the burst is attributed to the properties of the saddle-homoclinic bifurcation. Consequently, the bursting waveform shown in Fig. 9 (b) and the burst trajectory in Fig. 11 (b) are qualitatively similar to the ones shown in Fig. 2 (b) and Fig. 4 (b) for the $I_{Na,p}+I_K+I_{K(M)}$ model. This qualitative similarity is because the bursts are initiated and terminated through the same types of bifurcation in both systems.

### B. Subcritical Andronov-Hopf / Fold Limit Cycle Bifurcation

In this section, we discuss a modified version of the circuit in Fig. 8, where we only remove the leak channel represented by $R_3$ as shown in Fig. 12. We also adjust the parameters of the circuit such that the circuit outputs sustained bursting, where each burst is initiated via a subcritical Andronov-Hopf bifurcation and terminated via a fold limit cycle bifurcation. At $I = 5\ uA$, the circuit exhibits neither spiking nor bursting, and the membrane potential converges to a resting state as shown in Fig. 13 (a). However, when the injected current is increased to $I = 5.6\ uA$, the circuit undergoes a bifurcation that results in the transition to the bursting state shown in Fig. 13 (b).

As in the previous sections, we use the method of dissection of neural bursting to analyse the bursting behaviour of the circuit. We first obtain equations 10, 11, and 12, describing the evolution of the state variables of the circuit.

$$\frac{dV_{out}}{dt} = \frac{1}{C_1}\left(-\frac{V_{out}-V_{GS1}}{R_1} - i_{DSQ_1} + i_{DSQ_3} - \frac{V_{out}-V_{GS2}}{R_2} - i_{DSQ_4}\right) \quad (10)$$

$$\frac{dV_{GS1}}{dt} = \frac{1}{C_2 R_1}(V_{out}-V_{GS1}) \quad (11)$$

$$\frac{dV_{GS2}}{dt} = \frac{1}{C_3 R_2}(V_{out}-V_{GS2}) \quad (12)$$

Once again, we assume that the fast subsystem comprises the state variables $V_{out}$ and $V_{GS1}$, and the slow subsystem comprises only the state variable $V_{GS2}$. Treating $V_{GS2}$ as the bifurcation parameter, we plot the nullclines of the fast subsystem with varying $V_{GS2}$ in Fig. 14. The mechanism underlying bursting here is similar to the one discussed in the previous section. However, the type of bifurcation that causes the bursting is different, which results in different bursting characteristics. Fig. 14 (b) shows the nullclines of the fast subsystem of the circuit at $V_{GS2} = 0.62\ V$. The nullclines in Fig. 14 (b) show that the fast subsystem has a stable focal equilibrium, an unstable limit cycle, and a stable limit cycle.

The coexistence of the stable equilibrium and the stable limit cycle implies that the fast subsystem is bistable. Therefore, the membrane potential may reside at either the stable equilibrium or the stable limit cycle, depending on whether the circuit is in the quiescent or spiking phases of bursting. During the quiescent phase of bursting, $V_{out}$ is smaller than $V_{GS2}$, which causes $V_{GS2}$ to decrease towards $V_{out}$. Consequently, the unstable limit cycle shrinks around the stable equilibrium as shown in Fig. 14 (b). With further decrease in $V_{GS2}$, the unstable limit cycle disappears, and the focal equilibrium loses stability through a subcritical Andronov-Hopf bifurcation as shown in Fig. 14 (a). As a result, the membrane potential converges to the stable limit cycle, and a burst is initiated.

During the spiking phase of bursting, $V_{out}$ is larger than $V_{GS2}$, which causes $C_3$ to charge and $V_{GS2}$ to increase. Consequently, the focal equilibrium regains stability through the same subcritical Andronov-Hopf bifurcation, and the unstable limit cycle reappears. This unstable limit cycle separates the attraction domains of the stable equilibrium and the stable limit cycle. Thus, the membrane potential remains at the stable limit cycle. As $V_{GS2}$ increases further, the unstable limit cycle expands and approaches the stable limit cycle as shown in Fig. 14 (c). Eventually, the stable and unstable limit cycles coalesce and annihilate through a fold limit cycle bifurcation, leaving only the stable equilibrium as shown in Fig. 14 (d). Therefore, the membrane potential converges to the stable equilibrium, and the burst is terminated.

Since the resting value of the stable equilibrium is lower than $V_{GS2}$, $C_3$ starts to discharge and $V_{GS2}$ decreases. Consequently, the stable and unstable limit cycles reappear through the fold limit cycle bifurcation. However, the membrane potential remains at the attraction domain of the stable equilibrium until the equilibrium loses stability through the subcritical Andronov-Hopf bifurcation as $V_{GS2}$ decreases further. Once the equilibrium loses stability, the membrane potential converges to the stable limit cycle, and another burst is initiated. This process results in the sustained bursting shown in Fig. 13 (b) and is qualitatively analogous to the process governing the bursting of the $I_{Na,p}+I_K+I_{K(M)}$ model discussed in section II-B.

Fig. 15 (a) shows the bifurcation diagram of the fast subsystem with varying $V_{GS2}$. The bifurcation diagram illustrates that two bifurcations of the fast subsystem occur. The subcritical Andronov-Hopf bifurcation at $V_{GS2} \approx 0.615\ V$ is responsible for the initiation of the burst, which is attributed to the equilibrium's loss of stability. While the fold limit cycle bifurcation at $V_{GS2} \approx 0.659\ V$ is responsible for the termination of the burst, which is due to the disappearance of the stable limit cycle. For $V_{GS2} < 0.615\ V$, the fast subsystem has only an unstable equilibrium and a stable limit cycle. Therefore, the membrane potential always converges to the stable limit cycle, corresponding to the spiking phase. In contrast, for $V_{GS2} > 0.659\ V$, the fast subsystem has only a stable equilibrium. Thus, the membrane potential always converges to that stable equilibrium, which corresponds to the quiescent phase. However, for $0.615\ V < V_{GS2} < 0.659\ V$, both the stable equilibrium and the stable limit cycle coexist. Therefore, the fast subsystem becomes bistable and exhibits a hysteresis effect. This implies that the membrane potential would converge to either the stable limit cycle or the stable equilibrium depending on which attraction domain it is initially within. The bifurcation diagram in Fig. 15 (a) is qualitatively equivalent to that of the $I_{Na,p}+I_K+I_{K(M)}$ model in Fig. 7 (a), which highlights the qualitative similarity between the two systems.

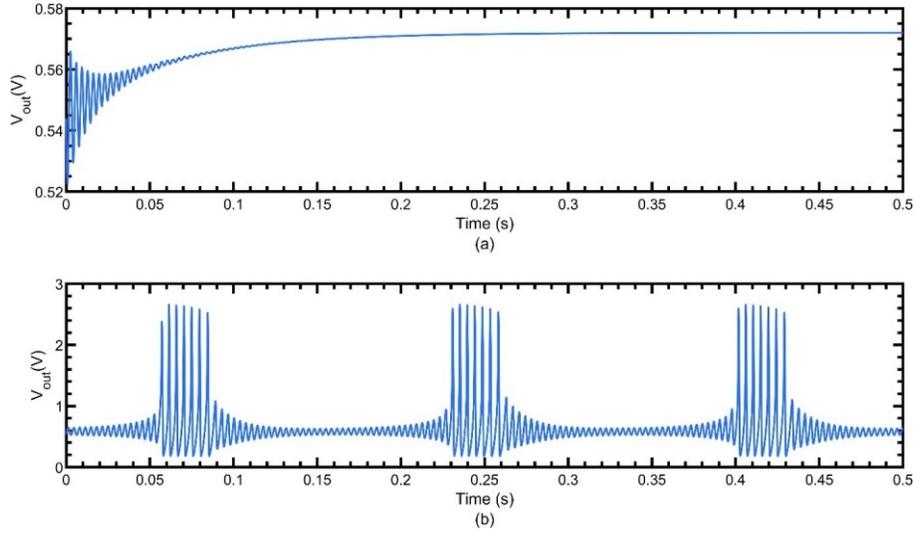

Fig. 13. Membrane potential output of the FET-based minimal bursting neuron circuit in Fig. 12 with $C_1 = 5\ nF$, $C_2 = 1\ nF$, $R_1 = 1\ M\Omega$, $V_{dc} = 3.5\ V$, $K_{n1} = 40\ uA/V^2$, $V_{to1} = -0.5\ V$, $\lambda_1 = 0.01$, $K_{p2} = 100\ uA/V^2$, $V_{to2} = 2\ V$, $\lambda_2 = 0.01$, $K_{n3} = 100\ uA/V^2$, $V_{to3} = -2\ V$, $\lambda_3 = 0.01$, $C_3 = 100\ nF$ $R_2 = 1\ M\Omega$, $K_{n4} = 40\ uA/V^2$, $V_{to4} = 0.3\ V$ and $\lambda_4 = 0.01$. (a) I = 5 uA. (b) I = 5.6 uA.

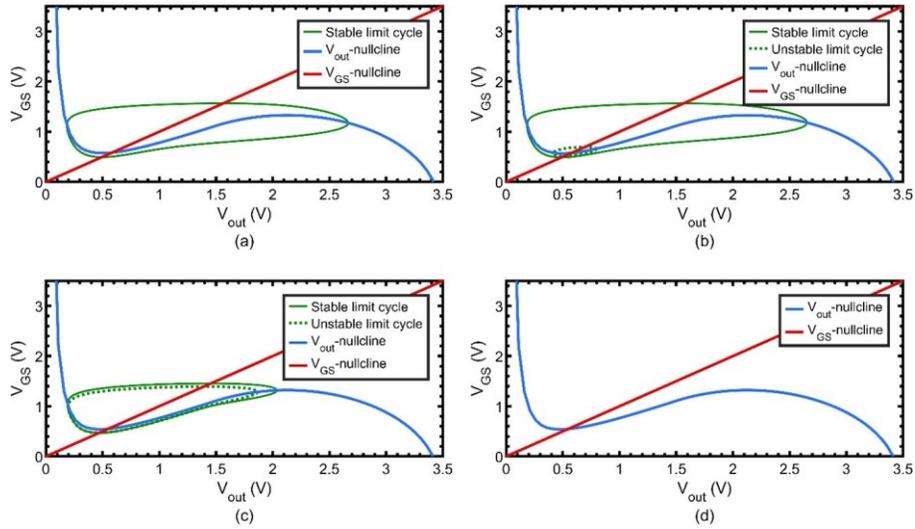

Fig. 14. Nullcline analysis of the FET-based minimal bursting neuron circuit in Fig. 12 with the same parameters values as Fig. 13 (b). (a) $V_{gs2}$=0.61 V. (b) $V_{gs2}$=0.62 V. (c) $V_{gs2}$=0.6583 V. (d) $V_{gs2}$=0.66 V.

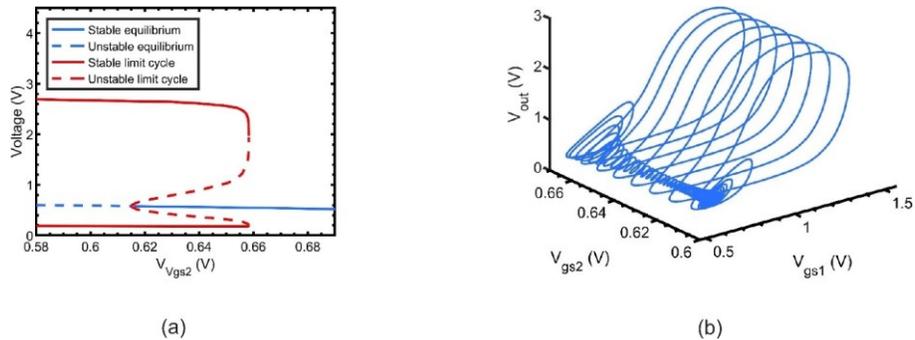

Fig. 15. (a) Bifurcation diagram and (b) trajectory of membrane potential bursting of the FET-based minimal bursting neuron circuit in Fig. 12. The same parameters values in Fig. 13 (b) were used.

TABLE I
COMPARISON OF SPIKING NEURON IMPLEMENTATIONS

| Neuron Circuits | Neuron Model | Number of Components | Signal Gain | Afterhyper-polarization | Subthreshold Oscillations | Biologically Plausible | I/O variables share same node | Bursting |
|---|---|---|---|---|---|---|---|---|
| Proposed circuit in Fig. 8 | $I_{Na,p}+I_K+I_{K(M)}$ | 4 FETs, 3 capacitors, 3 resistors | ✓ | ✓ | ✗ | ✓ | ✓ | ✓ |
| Proposed circuit in Fig. 12 | $I_{Na,p}+I_K+I_{K(M)}$ | 4 FETs, 3 capacitors, 2 resistors | ✓ | ✓ | ✓ | ✓ | ✓ | ✓ |
| [20] | Izhikevich | 14 FETs, 2 capacitors | ✓ | ✓ | ✗ | ✓ | ✓ | ✓ |
| [21] | HH | 6 FETs, 4 capacitors | ✓ | ✓ | ✗ | ✓ | ✓ | ✗ |
| [22] | HH | 4 FETs, 2 OTAs, 3 capacitors | ✓ | ✓ | ✗ | ✓ | ✓ | ✗ |
| [23] | LIF | 34 FETs, 1 capacitor | ✗ | ✗ | ✗ | ✗ | ✗ | ✗ |
| [24] | LIF | 20 FETs, 1 capacitor | ✓ | ✗ | ✗ | ✗ | ✗ | ✗ |
| [25] | Morris-Lecar | 8 FETs, 2 capacitors, 1 resistor | ✓ | ✓ | ✗ | ✓ | ✓ | ✗ |
| [26] | N/A | 12 FETs, 2 capacitors, 1 comparator | ✓ | ✗ | ✗ | ✗ | ✗ | ✓ |
| [27] | N/A | 2 NbO$_2$ Mott memristors, 3 capacitors, 2 resistors | ✓ | ✓ | ✗ | ✗ | ✗ | ✓ |
| [28] | N/A | 2 VO$_2$ Mott memristors, 2-3 capacitors, 1-2 resistors | ✓ | ✓ | ✓ | ✗ | ✗ | ✓ |

As mentioned in the previous sections, the type of bifurcation undergone at the start and end of each burst affects the characteristics of the resulting bursting. The subcritical Andronov-Hopf bifurcation at the start of each burst results in the equilibrium losing stability, but is not responsible for the appearance of the stable limit cycle. The stable limit cycle exists independently of the subcritical Andronov-Hopf bifurcation. Therefore, the spikes at the beginning of each burst commence with nonzero amplitude. The fold limit cycle bifurcation that terminates each burst does not cause the stable limit cycle to shrink, but rather to disappear abruptly. Consequently, the spikes at the end of each burst do not decay to a zero amplitude, but rather converge to the equilibrium abruptly. The bursting in Fig. 13 (b) exhibits subthreshold oscillations, which is further highlighted by the burst trajectory in Fig. 15 (b). The presence of subthreshold oscillations is due to the equilibrium being of the focal type. The characteristics of the bursting illustrated in Fig. 13 (b) are qualitatively similar to that of the $I_{Na,p}+I_K+I_{K(M)}$ model in Fig. 5 (b). This qualitative similarity is attributed to the fact that in both systems the bursts are initiated and terminated through the same types of bifurcation.

## IV. DISCUSSION

In this work, we proposed two novel minimal bursting neuron circuits. We demonstrated that the proposed circuits exhibit biologically plausible bursting with neuronal characteristics similar to that of the $I_{Na,p}+I_K+I_{K(M)}$ model. In this section, we compare our bursting neuron circuits to state-of-the-art circuits from the literature. Table I provides a summary of this comparison. The circuit in [20] implements the Izhikevich neuron model and is therefore capable of bursting. However, this circuit requires a significantly larger number of FETs than our proposed designs. The circuits in [21] and [22] implement the Hodgkin-Huxley model, thus, they are unable to burst. Notably, the circuit proposed in Fig. 12 achieves both subthreshold oscillations and bursting while using fewer components than [21] and [22].

The neuron circuits in [23] and [24] are implementations of the LIF neuron model, which means they lack biological plausibility and are incapable of bursting. Additionally, the implementations in [23] and [24] use a very large number of components in comparison to the two neuron circuits we propose in this work. The memristive neuron implementations reported in [27] and [28] are able to exhibit bursting, along with signal gain and hyperpolarization. However, as highlighted in [17], [18], [29], their qualitative characteristics significantly differ from established neuron models, making them not biologically plausible.

## V. CONCLUSION

Large-scale neuromorphic systems require neuron circuits that are both compact and capable of exhibiting rich neuronal dynamics. These requirements create a trade-off between scalability and biological plausibility. This work addresses this challenge by proposing a design methodology for compact bursting neuron circuits that emulate the $I_{Na,p}+I_K+I_{K(M)}$ model. The proposed methodology yields neuron circuits that are biologically plausible without sacrificing the potential for scalability. Consequently, the proposed circuits are promising candidates for use as neuron blocks in large-scale neuromorphic implementations such as the Neurogrid [30] and FACETs [31]. Moreover, the intrinsic bursting dynamics of these circuits enable SNN implementations that exploit resonance-based learning, allowing a presynaptic neuron to selectively target an individual postsynaptic neuron and thereby facilitate learning. Overall, this work establishes a foundation for the development of scalable and biologically plausible neuromorphic systems.